\documentclass[9pt,a4paper,twocolumn]{article}

\usepackage[utf8]{inputenc}
\usepackage{dcolumn}
\usepackage{bm}
\usepackage{braket}
\usepackage[T1]{fontenc} 
\usepackage{float}
\usepackage{amsmath,amsfonts,amssymb,eqnarray}  
\usepackage{color, soul}
\usepackage{graphicx}  
\usepackage{dblfloatfix} 
\usepackage{array,booktabs}
\usepackage[pdftex,hypertexnames=false]{hyperref} 
\usepackage[affil-it]{authblk}
\usepackage[font={small}]{caption}
\usepackage{cite}
\usepackage[left=2cm,right=2cm,top=1.6cm]{geometry}
\usepackage{titlesec}
\titleformat{\section}
  {\normalfont\fontsize{13}{15}\bfseries}{\thesection}{0.5em}{}
\titleformat{\subsection}
  {\normalfont\fontsize{11}{15}\bfseries}{\thesubsection}{0.5em}{}
\titleformat{\subsubsection}
  {\normalfont\fontsize{10}{15}\bfseries}{\thesubsubsection}{0.5em}{}  

\title{High-dimensional quantum communication: benefits, progress and future challenges}
\author{Daniele Cozzolino}
\author{Beatrice Da Lio}
\author{Davide Bacco*}
\author{Leif Katsuo Oxenløwe}
\affil{\small SPOC, DTU Fotonik, Technical University of Denmark, Kgs. Lyngby 2800, Denmark}
\affil{$^*$\small dabac@fotonik.dtu.dk}

\date{\today}

\begin{document}

\twocolumn[ 
\begin{@twocolumnfalse}
\maketitle
     \vspace{-0.8cm}
  \begin{abstract}
      \normalsize
         \vspace{1pt}
\noindent {In recent years, there has been a rising interest in high-dimensional quantum states and their impact on quantum communication. Indeed, the availability of an enlarged Hilbert space offers multiple advantages, from larger information capacity and increased noise resilience, to novel fundamental research possibilities in quantum physics. Multiple photonic degrees of freedom have been explored to generate high-dimensional quantum states, both with bulk optics and integrated photonics. Furthermore, these quantum states have been propagated through various channels, \textit{e.g.} free-space links, single-mode, multicore, and multimode fibers and also aquatic channels, experimentally demonstrating the theoretical advantages over two-dimensional systems.
Here, we review the state of the art on the generation, the propagation and the detection of high-dimensional quantum states.}
\end{abstract}
\vspace{.5cm}  
  \end{@twocolumnfalse}]

\section{Introduction}
The advent of quantum information has strongly influenced modern technological progress. Intense research activities have been carried out in the last two decades on such field, producing outstanding results, \textit{e.g.} in quantum computing~\cite{divincenzo1995quantum,vandersypen2001experimental,knill2001scheme,ronnow2014defining,boixo2018characterizing}, communication~\cite{yin2017satellite,bouwmeester1997experimental,ren2017ground,minder2019experimental} and simulation~\cite{lloyd1996universal,sparrow2018simulating,britton2012engineered,barreiro2011open,giordani2018}. A \textit{qubit}, the quantum counterpart of the classical bit, is a two-level quantum system and constitutes the elementary unit of quantum information. Qubit manipulation and control were demanding tasks at first, but are now routinely used in quantum experiments.
It is interesting to investigate quantum information in larger Hilbert spaces, either by increasing the number of qubits or by exploiting $d$-level quantum systems, that is \textit{qudits}.
Great advantages derive from accessing Hilbert spaces of higher dimensions, and whether it is better to increase the number of qubits or to exploit qudits only depends on the particular task to accomplish. In this review, we will discuss the advantages derived by using high-dimensional states, \textit{i.e.} qudits, focusing our attention on those related to quantum communication\footnote{Note that, throughout this review, entanglement among qudit will be discussed by considering only a maximum of 2 particles involved. Moreover, certification of high-dimensional entanglement is not discussed throughout the text, but an extensive review on certification methods and experiments is reported in Ref.~\cite{friis2018entanglement}}. Nonetheless, high-dimensional quantum states have shown to yield improvements in several other fields. Indeed, they allow to achieve increased sensitivity in quantum imaging schemes~\cite{lloyd2008enhanced}, they can boost the transport efficiency of biological compounds~\cite{sarovar2010quantum}, they constitute richer resources for quantum simulation~\cite{neeley2009emulation,kaltenbaek2010optical}, they lead to higher efficiencies in quantum computing~\cite{lanyon2009simplifying,bocharov2017factoring,babazadeh2017high,muralidharan2017overcoming} and clock synchronization~\cite{tavakoli2015quantum}, and they can be beneficial in quantum metrology applications~\cite{fickler2012quantum}. Moreover, breakthrough experiments studying quantum information memories have been performed by using
high-dimensional states~\cite{ding2016high,parigi2015storage,dkabrowski2018certification}.\\
Throughout the review, the physical $d$-level systems we are going to refer to are photons, but quantum information processing with other high-dimensional physical systems is also possible~\cite{kiktenko2015multilevel,kiktenko2015single}. In the first part, we discuss high-dimensional quantum states as a resource and we summarize the benefit derived by their exploitation. The second and third parts constitute the core of this work. In the former, we summarize experimental sources of qudits, either bulk or integrated. In the latter, we present a compendium on quantum communication experiments with high-dimensional states, organized according to the link adopted, that is free-space, fiber or underwater links. Finally, we highlight possible perspectives for high-dimensional quantum communication and raise several questions for future investigations.

\section{Enlarging Hilbert spaces: the more the better}
Qubits are the basic quantum information units and are described by a basis of two orthonormal vector states, $\lbrace \ket{0},\ket{1}\rbrace$, corresponding to the classical bits 0 and 1 respectively. Conversely, a qudit is a quantum system that is not constrained into a two-dimensional space and that, in principle, can have any integer number $d$ of levels. However, besides conceptual limitations-- \textit{e.g.}, how large has to be a quantum state before exiting the quantum realm?-- current experimental devices pose an upper bound on the number of dimensions that can be coherently controlled. In this section, we are going to highlight the advantages offered by such high-dimensional systems, presenting both theoretical and experimental results.

\subsection{Information and communication capacity}
The first and rather clear advantage offered by qudits is the increased information capacity per quantum system. For example, by using high-dimensional states with $d=4$ (\textit{ququarts}), 2 bits of information can be encoded: $\ket{0}=00$, $\ket{1}=01$, $\ket{2}=10$ and $\ket{3}=11$. A quantitative measure of the larger information capacity is given by the relation $\mathrm{log}_2d$, which returns the number of classical bits (or qubits) needed to encode the same amount of information~\cite{cover2012elements}. Additionally, high-dimensional entangled states yield a larger channel capacity, \textit{i.e.} the amount of information reliably transmitted over a communication channel. Entanglement was predicted by A.~Einstein, B.~Y.~Podolsky and N.~Rosen~\cite{epr1935} and causes quantum non-local correlations that cannot be devised by any local theory~\cite{bell1964}. In Refs.~\cite{barreiro2008beating,hu2018beating,dixon2012quantum}, it has been experimentally demonstrated how bipartite entangled qudit can beat the classical channel capacity, with (Refs.~\cite{barreiro2008beating,hu2018beating}) or without (Ref.~\cite{dixon2012quantum}) superdense coding schemes. Nonetheless, there exists a capacity limit for direct communication between two parties, the so-called PLOB-repeaterless bound~\cite{pirandola2017fundamental}. However, a recent work by D.~Miller~\cite{miller2019parameter} shows how the PLOB bound can be surpassed using error-corrected qudit repeaters, by analyzing different parameter regimes.

\subsection{Higher noise resilience}
Along with the increased information capacity, high-dimensional quantum states own a very important feature for quantum communication, that is they are more robust to noise, either if it is environmental or derived from eavesdropping attacks. 
\begin{figure}[ht!]
\centering
\includegraphics[width=0.53\textwidth]{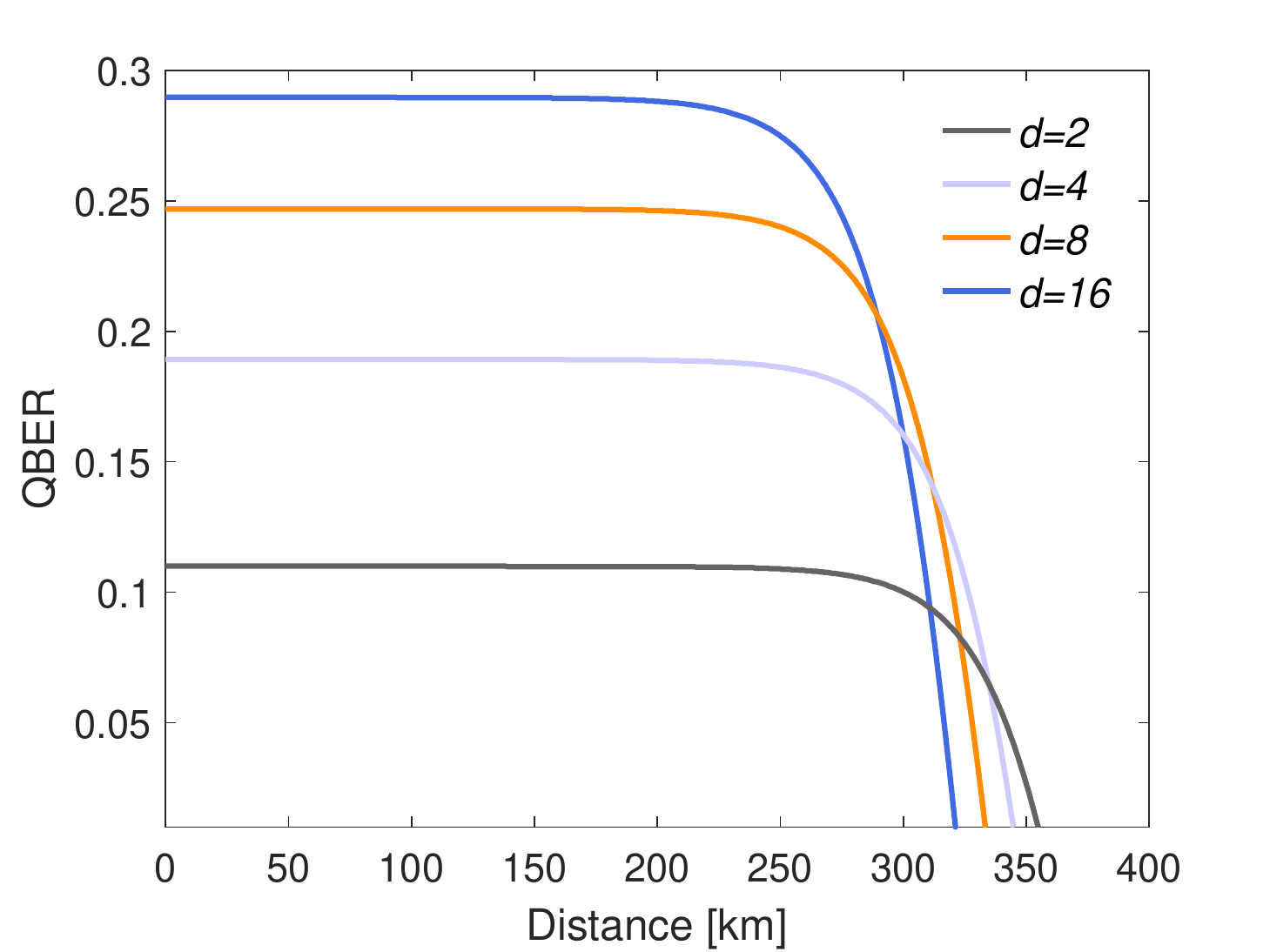}%
\caption{{\bf Maximum error tolerance for a positive secret key rate as a function of the distance and different dimensions.} The curves have been derived by considering a single-photon $d$-dimensional BB84 protocol, using ideal detectors affected only by the dark count probability ($p_d=10^{-8}$ per detector) and in case of coherent attacks~\cite{sheridan2010security,cerf2002security}. As attenuation, we consider the standard single-mode fiber parameter $\alpha=0.2$ dB/km and we assumed to have $d$ detectors to measure $d$ states simultaneously. Each curve identifies a region within which a positive secret key rate can be extracted. The maximum attainable transmission distance decreases by increasing
the state dimension $d$, indeed the greater $d$ is the more the states are sensitive to the dark counts of the detectors.}
\label{fig:maxqber}
\end{figure}
Indeed, the security of a quantum channel, which is guaranteed by quantum physical laws, is the cornerstone for sharing encrypted random keys (quantum key distribution-- QKD), but also for general quantum communication protocols. The security of a settled quantum link is ensured by having the quantum bit error rate (QBER), \textit{i.e.} the ratio of an error rate to the overall received rate, below a certain threshold. In the case of qubit-based protocols, the threshold value has been proven to be 11\% against the more general coherent attacks and by using two mutually unbiased bases (MUBs) in one-way reconciliation~\cite{scarani2009security}. The higher resilience to noise sources owned by qudits has been shown in Refs.~\cite{sheridan2010security,cerf2002security,bechmann2000quantum}, where the information gained by a potential eavesdropper, Eve, performing coherent attacks is calculated both considering the use of $2$ and $d+1$ MUBs. As a result, it has been demonstrated that the robustness to noise of qudits increases with their dimension $d$, that is QBERs threshold values that ensure secure communication increase. For instance, for $d=4$ and $d=8$, the thresholds are 18.93\% and 24.70\% respectively, by using $2$ MUBs~\cite{sheridan2010security,cerf2002security}.
Such higher noise tolerance has also implications on the final secret key rate. Indeed, for fixed noise level the secret key rate increase with the Hilbert space dimensions. In Figure~\ref{fig:maxqber}, we show the maximum acceptable error rate to generate a positive secret key rate as a function of the distance and for different qudit dimensions. The curves are derived from Refs.~\cite{sheridan2010security,cerf2002security} and refer to an ideal system performing a single-photon $d$-dimensional BB84 protocol~\cite{bb84}, using ideal detectors affected only by the dark count probability ($p_d$) and assuming coherent attacks. For each dimension, a region within which a positive secret key rate can be extracted is identified. Besides, the achievable transmission distance decreases by increasing dimensions and suggests qubit protocols to reach the longest distance. However, in a practical scenario, the actual advantages of high-dimensional states over qubits strongly depend on the particular physical implementation, which varies the operational constraints required. Thus, there might be cases where high-dimensional states perform better than qubits also in terms of transmission distance~\cite{Cozzolino2018,Bacco2017}.\\
The higher noise resilience of qudits also has advantages if they are entangled. Indeed, it has been demonstrated that the security of the E91 protocol~\cite{e91}, generalized to the qudit case, is ensured with increased error thresholds if the dimensions are increased as well. Such a conclusion implies that the robustness of the quantum correlations is influenced by the dimension of the Hilbert space. In Ref.~\cite{e91} this result has been theoretically proven. In particular, the authors analyze how the number of dimensions $d$ and the number of entangled particles $N$ affect the entanglement robustness for a phase damping and depolarizing channel. Although one might expect that the dependence of the entanglement robustness on $N$ and $d$ could be similar (in both cases the overall Hilbert space is enlarged), they show that the entanglement becomes more fragile by increasing $N$ (with $d$ fixed), whereas it is more robust by increasing $d$ (with $N$ fixed)~\cite{liu2009decay}. This conclusion can be intuitively understood by considering an example. If the dimension $d$ is fixed, by increasing $N$, the components of the state, \textit{i.e.} the number of photons constituting the final system $\rho$, are increasing. During the transmission through a channel, the noise sources will act locally on every system, thus making the entanglement more fragile with the growth of $N$. Conversely, if $N$ is fixed, the influence of the noise sources in the channel on $\rho$ will be more and more negligible by increasing $d$, which leads to a more robust entanglement~\cite{liu2009decay}. Hitherto, the only experiment demonstrating the robustness of high-dimensional entangled states has been carried out in Vienna by S.~Ecker and co-authors~\cite{ecker2019entanglement}. They certified entanglement either with time-energy or with photonic orbital angular momentum (OAM) degrees of freedom (see Section 3) in different noise conditions, by changing the noise level during the measurements. As theoretically predicted, in both cases the noise level threshold increases with the qudit dimension.
In particular, time-energy entangled pairs with dimension $d=10$ can tolerate $56\%$ of noise in the channel before the entanglement breaks and such threshold increases to $93\%$ when $d=80$. For photon pairs entangled in the OAM degree of freedom, instead, entanglement breaks with $38\%$ of noise in the channel when the system has dimension $d=2$, whereas with $d=7$ value of $73\%$ of noise can be tolerated~\cite{ecker2019entanglement}.

\subsection{Enhanced robustness to quantum cloning}
The backbone for the security of quantum communication protocols is the no-cloning theorem, which states that an unknown quantum state cannot be perfectly copied~\cite{wootters1982single}. Although creating a perfect copy of an unknown quantum state is forbidden, it is possible to make imperfect clones, each with fidelity-- \textit{i.e.} the overlap between the initial state to be cloned and the cloned copy-- lower than one, where one corresponds to the perfect determination of the initial state~\cite{buvzek1996quantum,scarani2005quantum}. If the cloning scheme maximizes the attainable fidelity of the copied state, it is called optimal quantum cloning and if it does not depend on the initial state, it is said to be universal~\cite{gisin1997optimal,bruss1998optimal,ghosh1999optimal}. The most common cloning scheme produces cloned systems- \textit{i.e.}, all output states of the cloning machine- characterized by the same cloned state and thus is called \textit{symmetric}~\cite{gisin1997optimal,bruss1998optimal}. Nonetheless, \textit{asymmetric} quantum cloning is also possible~\cite{cerf2000asymmetric}. For the symmetric cloning, given $N$ copies of the initial state and $M>N$ imperfect cloned copies ($N\rightarrow M$), the optimal cloning fidelity in a $d$-dimensional Hilbert space is given by:
\begin{equation}
F^d_{clon}(N,M)=\frac{M-N+N(M+d)}{M(N+d)}    
\end{equation}
which reduces to $F^d_{clon}=\frac{1}{2}+\frac{1}{1+d}$ in the case $1\rightarrow 2$, that is one input state and two imperfect copies~\cite{navez2003cloning,bruss1999optimal}. Thus, by increasing the input state dimension, the cloning fidelity decreases from the upper bound $F^2_{clon}=0.83$ for qubits to $F^\infty_{clon}=0.50$ when $d$ approaches infinity. Such a feature clearly shows the benefit of high-dimensional states for quantum cryptography.\\
To our knowledge, only two experiments testing optimal quantum cloning fidelities with qudits have been performed~\cite{nagali2010experimental,bouchard2017high}. They are both based on the symmetrization method described in Refs.~\cite{sciarrino2004realization,ricci2004teleportation,irvine2004optimal}. In Ref.~\cite{nagali2010experimental} an optimal quantum cloning $1\rightarrow 2$ is carried out for ququarts, \textit{i.e.} four-dimensional states, encoded in the polarization and OAM degrees of freedom of a single-photon. In Ref.~\cite{bouchard2017high} instead, $1\rightarrow 2$ universal optimal quantum cloning is performed for arbitrary input states with dimensions up to $d=7$ and, to show the enhanced robustness of the high-dimensional states, cloning attacks are performed on the BB84 protocol in $d=7$ dimensions.   

\subsection{Larger violation of local theories}
In 1935, the main concern Einstein, Podolsky and Rosen had on quantum mechanics was the violation of the local realism principle, which the three scientists considered necessary~\cite{epr1935}. Local realism assumes that every object has physical properties that are prior to and independent from any possible measurement carried out by an observer and that the causality is bounded by the speed of light, \textit{i.e.} the special relativity holds. However, in 1964, in his seminal work, John Bell showed how specific quantum measurements on qubits could not be explained by any local theory~\cite{bell1964}. Thus, a violation of the notorious Bell's inequality implies the impossibility to explain with local theories the correlations under investigation. Many experiments obtaining such violation have been carried out and, very recently, also in a loophole-free manner~\cite{Hensen2015,Shalm2015,Rosenfeld2017}. The generalization of the Bell's inequality to a system with higher dimensions was firstly studied by the pioneering works of N.~D.~Mermin and A.~Grag~\cite{Mermin1980,Grag1982}, and a few years later by A.~Peres~\cite{Peres1992}. In the early 2000s the advantages of high-dimensional states for violation of local-realism were shown by D.~Kaszlikowski \textit{et al.} and T.~Durt \textit{et al.}~\cite{Kaszlikowski2000,Durt2001}. Indeed they showed, with $3\leq d\leq 9$ first~\cite{Kaszlikowski2000} and with $d$ up to $16$ later~\cite{Durt2001}, how the violation increases with the dimensions, indicating enhanced robustness of the violation against the noise. Collins \textit{et al.}~\cite{cglmp2002} generalized the concept for a system of any dimensions $d$, obtaining an inequality often referred to as CGLMP inequality. Moreover, a very important result for fundamental research is reported in Ref.~\cite{vertesi2010}. Indeed, Vértesi and co-authors showed that, due to the higher local realism violation given by qudits, the detector efficiency required to close the detection loophole decreases with dimensions. Very recently, W.~Weiss \textit{et al.} challenged the robustness of high-dimensional states to violate Bell-like inequalities in a more practical scenario~\cite{Weiss_2016}. In particular, they studied the impact of imperfections in state-preparation and measurement settings on the violation of generalized nonlocality tests. Interestingly, these imperfections affect the violation in a dimension-dependent manner. Thus, it is possible to identify noise thresholds, for each dimension $d$, such that if exceeded the quantum-to-classical transition will emerge, making large systems behaving classically under Bell-like tests. Experimental violation of generalized Bell inequalities have been achieved with energy-time entangled qutrits ($d=3$ qudit)~\cite{Thew2004,Schwarz2014}, with entangled radial and angular degrees of freedom of Laguerre-Gauss (LG) modes for qudits with dimension $2\leq d \leq 10$~\cite{liu2019violation} and up to $d=12$ with entangled qudits encoded in the OAM degree of freedom~\cite{Dada2011}. The larger violation of Bell inequalities gives also benefit on entanglement-based device-independent QKD protocols. Indeed, to establish secure quantum key distribution, the randomness of the Bell measurements is needed. However, if the randomness is weak, above a certain threshold the communication cannot be considered secure. In Ref.~\cite{huberweak2013}, the authors show how the acceptable loss of randomness is significantly larger for qudit systems.\\
Performing typical communication protocols, like quantum teleportation or entanglement swapping, using high-dimensional states is still an open experimental challenge since they require a complete Bell state measurement. Indeed, it has been demonstrated in Refs.~\cite{Calsamiglia2002,Lutkenhaus1999,Vaidman1999,Michler1996} that projection onto a high-dimensional Bell state, such that it would be unequivocally identified, are unattainable with only linear optics elements. One way to work around this limitation is to use ancillary photons. Two experiments proved the principle very lately~\cite{hu2019experimental,luo2019quantum}. In Ref.~\cite{hu2019experimental} two auxiliary entangled photons are exploited to carry out a deterministic three-dimensional Bell state measurement, obtaining teleportation fidelities above 0.63 (surpassing the classical limit 0.5). In Ref.~\cite{luo2019quantum}, the authors draw a scheme where high-dimensional quantum teleportation can be realized by using $d-2$ ancillary photons, with $d$ being the Hilbert space dimension of the photon pair to be teleported. In their work, they realize qutrit teleportation by using only one ancillary photon, yielding a teleportation fidelity of 0.75.

\subsection{Advantages in communication complexity problems}
Communication complexity addresses problems on the amount of information that distributed parties need to share to accomplish a specific task~\cite{ccp,kumar2018experimental}. For instance, two parties, Alice and Bob, receive two inputs $x$ and $y$ respectively. They have to evaluate a certain function $f(x,y)$ without knowing which data the partner received. To improve the success of the protocol, before starting it, they are allowed to share classically correlated random strings or any other local data. Communication complexity problems can be mainly divided into two branches: one investigates the amount of communication required to all the possible parties to determine with certainty the value of $f(x,y)$; the other investigate which is the highest probability that the parties get to the correct value of $f(x,y)$ if only a limited amount of communication is allowed. From a quantum information perspective, the question that arises is whether there are some advantages in terms of complexity by using quantum correlations, that is entangled resources, in place of the classically correlated data. The issue has been studied by the pioneering works of H.~Buhrman, H.~Cleve and W.~Van Dam ~\cite{Buhrman2001,Cleve1997}, who showed that the highest success probability to determine $f(x,y)$ is given by $P_C=0.75$ by using classically correlated resources and $P_Q=0.85$ if Alice and Bob share a maximally entangled pair of qubits. Thus, in a classical protocol 3 bits of information are needed to compute $f$ at least with the probability $P_C=P_Q$, whereas 2 bits are sufficient if the protocol is supported by non-classical correlations~\cite{Buhrman2001}. Later on, this result has been generalized by \v{C}.~Brukner \textit{et al.}~\cite{Brukner2004,Brukner2002}. They showed that for every Bell's inequality, also in the case of high-dimensional systems, there always exists a communication complexity problem for which an entangled assisted protocol is more efficient than any classical one.\\
A second approach to deal with communication complexity problems is by using quantum communication instead of classical communication. Indeed, many communication tasks can be successfully performed, outperforming classical constraints in terms of communication complexity problems, either by entangled assisted classical protocol or by sending single quantum systems~\cite{guerin2016,Smania2016,galv2001,Casaccino2008}. In particular, experimental evidences in which the former performs better than the latter~\cite{paw2010,Hameedi2017} and vice-versa~\cite{Tavakoli2017,Tavakoli2016} can be found. However, D.~Mart\'inez \textit{et al.} have shown, both theoretically and experimentally, that high-dimensional quantum communication outperforms classical protocols assisted by nonlocal correlation whenever dimensions are $d\geq 6$~\cite{martinez2018}. Indeed, they demonstrated how dimension six acts as a threshold to reveal the benefits of quantum communication over implementations based on the violations of CGLMP inequalities. For dimensions below six, both communication complexity problem strategies are equally efficient, whereas for $d\geq 6$ they are not. Experimentally, they proved the statement by implementing qudits encoded in the linear transverse momentum of single-photons up to dimension $d=10$. Very recently, W.~Kejin \textit{et al.} demonstrated a communication complexity advantage given by high-dimensional protocols based on the quantum switch (a novel quantum resource which creates a coherent superposition of the causal order of events) for casually ordered protocols~\cite{switchcomplexity}.

\section{Qudit generation: how to expand Hilbert spaces}
After having declaimed the advantages gained in quantum communication by using larger Hilbert spaces, a question naturally arises: how can a qudit be practically implemented? The goal is to increase the available dimensions to send more than 1 bit/photon from one party (Alice) to another (Bob). To expand the Hilbert space, different photonic degrees of freedom or combination of them can be used. Within this section, we are going to discuss techniques and methods adopted to control those degrees of freedom and generate qudits.

\subsection{Bulk optics schemes}
\begin{figure*}[ht!]
\centering
\includegraphics[width=0.9\textwidth]{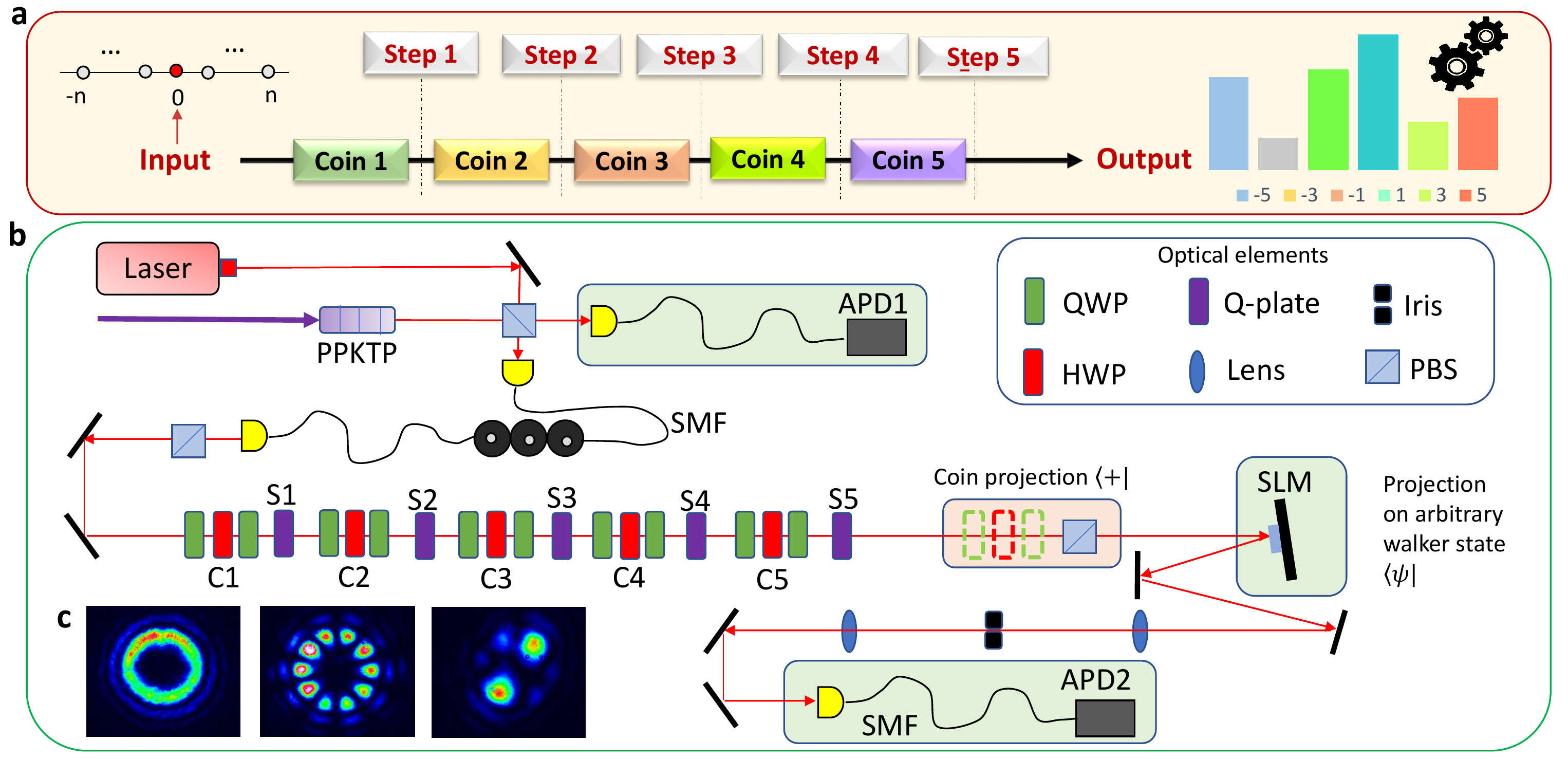}
\caption{{\bf Setup of the arbitrary qudits generation through quantum walks.} (a) Schematic of the protocol: at each step, the coin operator is changed to have the desired walker state at the output. (b) Experimental setup. A ppKTP source generates pairs of photons, which are then coupled to SMF. One photon acts as a trigger, while the other one is prepared in the initial $\ket{\psi_0}$ with a PBS and a polarization controller. Coin and shift operators are implemented with a set of wave plates and $q$-plates respectively. The detection consists of a PBS, followed by an SLM, an SMF and avalanche photodiode detectors (APD). (c) Pictures of the OAM modes after the PBS for (from right to left) OAM eigenstate
corresponding to $\ell=5$; balanced superposition of $\ell=\pm 5$; balanced superposition of all OAM components involved in the quantum walk dynamics, \textit{i.e.} $\ell=\pm 5, \pm 3, \pm 1$. Reprinted figure with permission from T.~Giordani \textit{et al.}, Physical review letters, vol. 122, no. 2, p. 020503, 2019 [107]. Copyright (2019) by the American Physical Society.}
\label{fig:quditRome}
\end{figure*}
\textit{Orbital angular momentum} -- The orbital angular momentum (OAM) of light is one of the most frequently exploited photonic properties to generate high-dimensional quantum states. Indeed, photons carrying an OAM different from zero are characterized by a helical phase factor $\mathrm{e}^{i\ell\phi}$, with $\phi$ being the azimuthal angle and $\ell$ the quantum number indicating the amount of orbital angular momentum $\ell\hbar$ carried by them. Since $\ell$ takes integer values and is unbounded, an arbitrarily large Hilbert space can be spanned. Thus, by properly controlling $\ell$, OAM offers a discrete basis to devise high-dimensional states. The optical fields describing such photons present a topological phase singularity at the beam axis, resulting in a characteristic ring-shaped intensity pattern in the case of classical light. Optical modes like LG modes~\cite{allen1992oam} or circular beams~\cite{vallone2015properties} carry a non-zero OAM, so progress on its manipulation is inevitably related to technological advances on waveshaping devices.
Several apparatuses, like cylindrical lenses~\cite{beijersbergen1993astigmatic}, specially designed laser cavities~\cite{tamm1990bistability}, spiral phase plates~\cite{beijersbergen1994helical} or integrated devices~\cite{cai2012integrated,chen2018mapping}, can be used to reshape the wavefront of an initial Gaussian photon and thus creating qudits encoded in OAM. More frequently experimental realizations make use of two other devices: holograms or $q$-plates~\cite{Rubano19,qplate2006}. Holograms can be considered as diffracting gratings such that the first-order diffracted beam acquires a unique phase and amplitude pattern. Holographic patterns can be easily created with commercially available devices called spatial light modulators (SLMs). A very different concept underlies $q$-plates, which apply suitable transformations on the local polarization state of light to generate phase shifts~\cite{qplate2006}. They are made of liquid crystals having an azimuthal pattern around a central point and confined within two slabs. The topological charge $q$ of the central singularity adds an OAM of $2\hbar q$ per photon and it can be an integer or half-integer. $Q$-plates are responsible for the spin-to-orbital angular momentum conversion, namely the exact conversion within the same quantum system of the spin angular momentum (SAM), \textit{i.e.} polarization, into OAM. Many experiments have been implemented either with SLMs or $q$-plates achieving quite important results~\cite{forbes2016creation,mair2001entanglement,fickler2012quantum,Vallone2014,d2012complete,marrucci2011spin}. Notwithstanding, it is worth mentioning for our purposes the experiment carried out recently in Rome by T.~Giordani \textit{et al.}~\cite{giordani2019experimental}, which combines both SLM and $q$-plates to generate arbitrary qudits through quantum walks~\cite{cardano2015quantum}. Indeed, by controlling the walk's dynamics through convenient step-dependent \textit{coin} operations, it is possible to steer the state of the \textit{walker} towards the desired final state. The experimental setup built for such a task is reported in Figure~\ref{fig:quditRome}. They implemented a discrete-time quantum walk with $n=5$ steps by using the OAM degree of freedom with $\ell=\pm 5, \pm 3, \pm 1$ to encode the walker state and the circular-polarization states $\lbrace \ket{R},\ket{L}\rbrace$ for the coin state. Five sets of half- and quarter-wave plates (HWP, QWP) are used to perform arbitrary coin operators, while five $q$-plates are used to implement the shift operator, which moves the walker conditionally to the
coin state. A type-II periodically poled potassium titanyl phosphate (ppKTP) crystal generates photons via parametric down-conversion. The two photons emitted are separated by a polarizing beam-splitter (PBS) and then coupled to a single-mode fiber (SMF): only one photon undergoes the quantum walk dynamics, whereas the other acts as a trigger. A first PBS sets the initial state of the walker and coin as $\ket{\psi_0}=\ket{0}_w\otimes\ket{+}_c$. The protocol requires a final projection on the state $\ket{+}_c$ of the coin, which can be performed by a final PBS. The OAM analysis is carried out by an SLM, allowing for an arbitrary OAM superposition detection. Five classes of qudit have been implemented: superposition of large OAM states, spin-coherent states, balanced states forming computational and Fourier bases, and random states. The average quantum fidelity obtained is $\bar{\mathcal{F}}=0.954\pm 0.001$, showing the correct implementation of all the desired multilevel quantum states.\\
Another very common method to generate photons with non-zero OAM is to get photons directly from a spontaneous parametric down-conversion (SPDC) process in $\chi^{(2)}$ materials, \textit{e.g.} ppKTP or $\beta$-barium borate (BBO) crystals. This technique is very suitable if photon pairs entangled in the OAM are intended to be used~\cite{mair2001entanglement}. Indeed, the conservation of the angular momentum in an SPDC process $\hbar\ell_p=\hbar\ell_1+\hbar\ell_2$ implies the generation of photon pairs with opposite OAM quantum numbers $\hbar\ell_1=-\hbar\ell_2$, if the pumping photons have $\hbar\ell_p=0$, that is if they are in a Gaussian mode. Thus, the theoretical states produced by the SPDC process are expressed as:
\begin{equation}
\ket{\Psi}=\sum_{\ell=-\infty}^{+\infty}c_\ell\ket{\ell}_1\ket{-\ell}_2
\label{spdc:oam}
\end{equation}
where $\ket{\pm\ell}_i$ are the photon states with OAM $\pm\ell$ and $c_\ell$ are complex probability amplitudes. Experimental conditions impose boundaries on $\ell$, \textit{i.e.} on the Hilbert space spanned, so that $\ell\in\lbrace -d,\dots,d\rbrace$. Suitable engineering of such photon sources allows the generation of very interesting states as four-dimensional Bell states~\cite{wang2017generation}, qutrits Greenberger-Horne-Zeilinger (GHZ) states~\cite{erhard2018experimental} or multi-photon entanglement in high dimensions~\cite{malik2016multi}. In Ref.~\cite{zhang2016engineering}, a very novel approach to generate OAM high-dimensional states is proposed. The authors carefully design an experimental setup in such a way that high-dimensional entangled states could be post-selected through Hong-Ou-Mandel interference~\cite{hom1987}.\\
\textit{Time} -- The process of SPDC allows the generation of qudits by also considering the time of photons emission as a degree of freedom. In this way, time-energy and time-bin entangled qudits can be produced. In the former case, the emission times of the photon pairs are undetermined with an uncertainty $\Delta t$ given by the Heisenberg uncertainty relation. The uncertainty $\Delta t$ can be described in terms of the coherence time of the pumping laser $t_{p}$, which is inversely related to the respective linewidth in the spectral domain. So, by pumping a crystal with a narrowband laser, longer coherence time is achieved, thus allowing for an increased uncertainty of the emission time of the photon pairs and giving rise to entanglement in the time-domain. The states generated by this process can be written as:
\begin{equation}
\ket{\Psi}=\sum_{k=1}^{d}\alpha_k\ket{k}_1\ket{k}_2
\label{spdc:time}
\end{equation}
where $\ket{k}$ refers to a photon in the $k$-th time slot within the coherence time of the pump and $\alpha_i$ is a complex probability amplitude~\cite{ecker2019entanglement}. Entangled qudits generated in this way can be measured by using unbalanced Mach-Zehnder interferometers, also known as Franson interferometers~\cite{franson1989bell}.\\ 
The generation of a time-bin entangled qudit differs only slightly from the time-energy one. Indeed, in the latter case, the discretization in different time states has to happen within the coherence length, whereas in the former this condition is not required.
\begin{figure}[t!]
\centering
\includegraphics[width=0.4\textwidth]{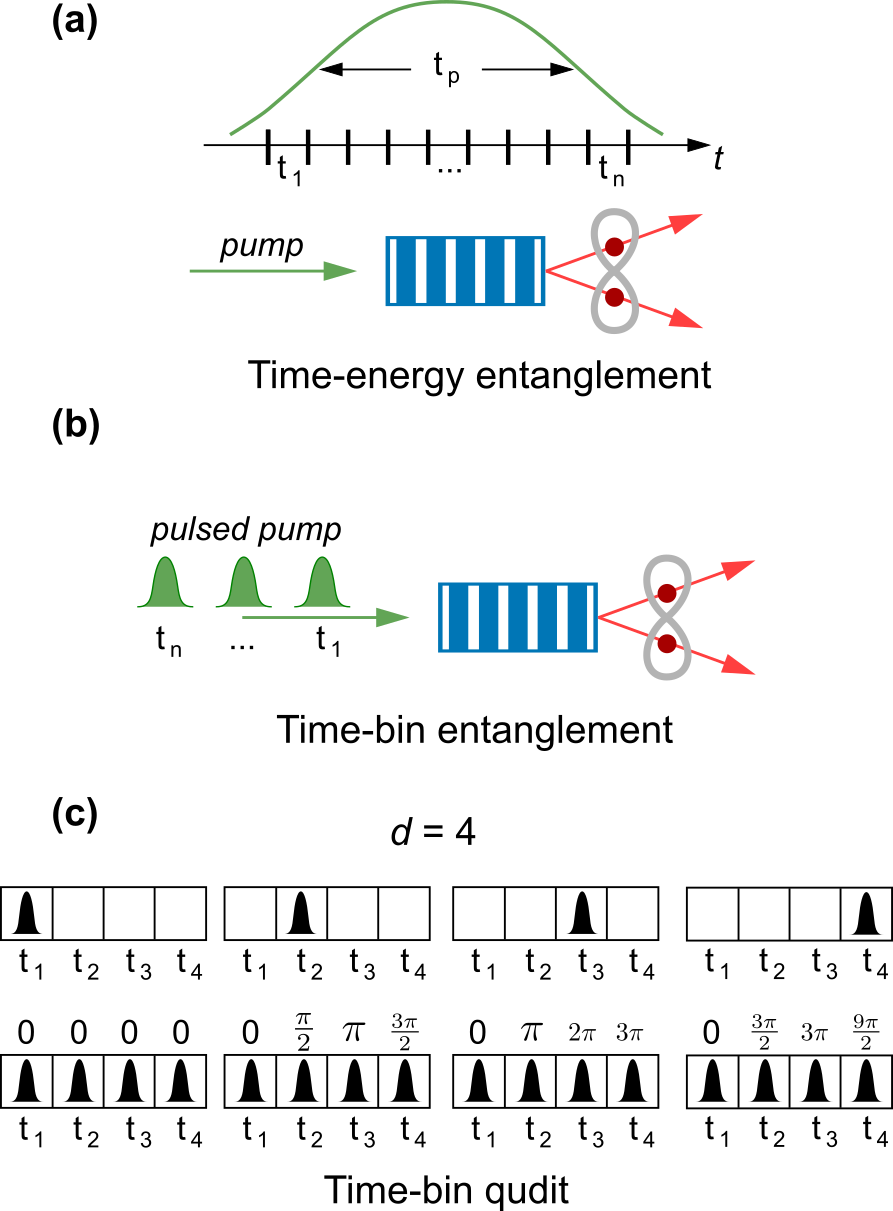}
\caption{{\bf Time encoding schemes.} (a) Time-energy entanglement. Within the coherence time of the pump $t_p$, photon pairs are generated with a temporal uncertainty $\Delta t$. Thus, by pumping with narrowband light, and in turn widening the coherence time, it is possible to define time slots, $t_1,\dots,t_n$, in which the photon pairs are generated. The number of time slots define the Hilbert space dimension and can be scanned with Franson interferometers. (b) Time-bin entanglement. In this case, the pump is pulsed and each pulse corresponds to a different time slot. (c) Time-bin qudit. Most frequently prepared by using attenuated pulses, time-bin qudits are very easily generated with off-the-shelf equipment as intensity and phase modulators. In terms of bases, the states of the computational one are identified by the pulse position, while those in the Fourier basis are identified by the relative phases among the pulses.}
\label{fig:timenc}
\end{figure}
\begin{figure*}[b]
\centering
\includegraphics[width=0.99\textwidth]{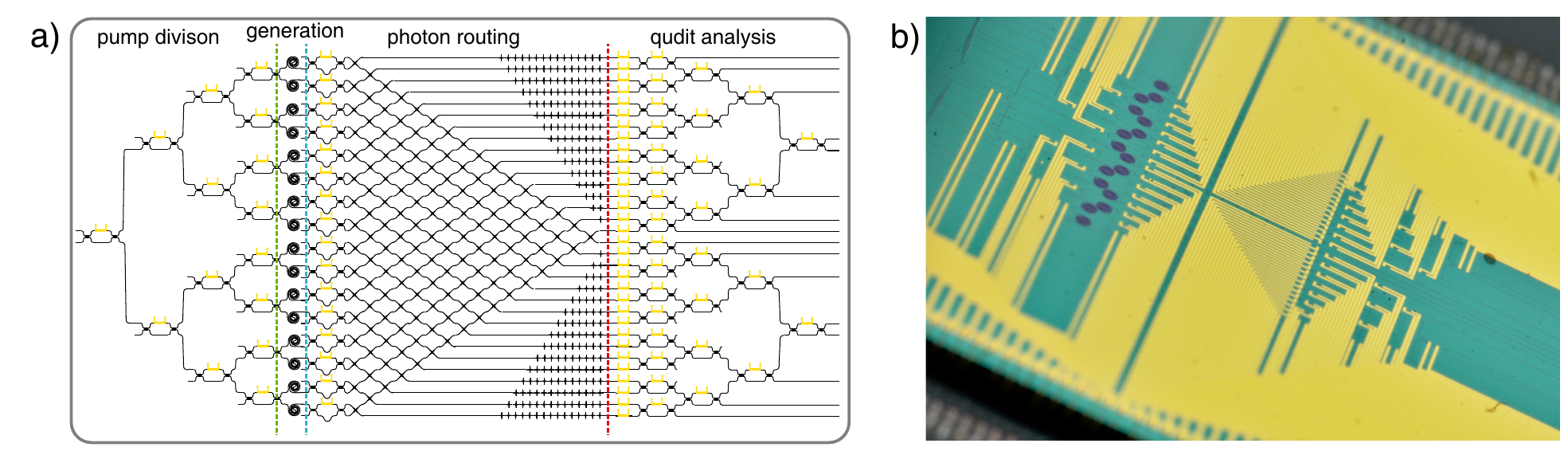}
\caption{{\bf Diagram and picture of the PIC.}
(a) Circuit diagram. By coherently pumping all the 16 sources a photon pair is generated in superposition across 16 optical modes, producing a multidimensional bipartite entangled state. The two photons, signal and idler, are routed through the chip by using asymmetric Mach-Zehnder interferometer (MZI) filters. By using triangular networks of MZIs, arbitrary local projective measurements are feasible. Photons are coupled off-chip and detected by two superconducting nanowire detectors. (b) Picture of the device.}
\label{fig:chip_bristol}
\end{figure*}
To generate time-bin entangled states, a pulsed pump is needed, so that each time state is identified by the pulse responsible for the photon pair generation~\cite{Marcikic2002}. The higher the number of pump pulses taken into account, the larger is the Hilbert space spanned. In this case, as before, Franson interferometers or combination of them are needed to reveal the entanglement among the states.\\
Interestingly, time-bin qudits can be easily created by using attenuated laser pulses (also called weak coherent pulses -- WCPs). Indeed, standard off-the-shelf equipment is required to generate them, making them widely adopted especially for QKD. Pulses in different time slots form the computational basis and, as before, the higher the number of time slots, the higher the dimension $d$ is. The superposition among these states is devised by controlling the relative phase of the pulses. Figure~\ref{fig:timenc} summarizes the time-encoding schemes we have discussed. Time encoding has been exploited in a large number of experiments, mostly focused on quantum communication and quantum cryptography~\cite{ecker2019entanglement,Steinlechner2017,Marcikic2002,Martin2017,Islam2017}.\\
\textit{Frequency} -- Generating high-dimensional quantum states in the frequency domain is also possible, although demanding with bulk optics instruments. Indeed, to our knowledge, not many experiments have been carried out, but many times this degree of freedom has been used in integrated photonic devices, as we are going to see in the following. Frequency entangled qudits can be generated starting from a parametric down-conversion process. The two-photon state from an SPDC process can be described as:
\begin{equation}
\ket{\Psi}=\int_{-\infty}^{+\infty}\mathrm{d}_s\mathrm{d}_if(\omega_s,\omega_i)\Hat{a}^\dagger_s(\omega_s)\Hat{a}^\dagger_i(\omega_s)\ket{0}
\label{spdc:freq}
\end{equation}
with $\Hat{a}^\dagger(\omega)$ being the creation operator at angular frequency $\omega$, the subscripts $s$ and $i$ indicating the signal and idler photons, respectively, and $f(\omega_s,\omega_i)$ being their joint spectral amplitude (JSA), which depends on the crystal and the pumping light. A clever way to create frequency entangled qudits is proposed in Ref.~\cite{Jin_2016}. The authors combined the photon pair state as described in eq. \eqref{spdc:freq} together with the Hong-Ou-Mandel interference. The obtained final state is expressed by: 
\begin{equation}
\begin{split}
\ket{\Phi(\tau)}=\frac{1}{\sqrt{\mathcal{N}}}\int_{-\infty}^{+\infty}\mathrm{d}_s\mathrm{d}_ih(\omega_s +\omega_i-\omega_p)\\
\times\delta(\omega_s +\omega_i-\omega_p)(1-\mathrm{e}^{-i(\omega_s-\omega_i)\tau})\\
\times\Hat{a}^\dagger_s(\omega_s)\Hat{a}^\dagger_i(\omega_s)\ket{0}    
\end{split} 
\label{qudit:freq}
\end{equation}
where $\mathcal{N}$ is a normalization factor, $\tau$ is the adjustable time delay between the two photons, $h(x)$ is a function dependent on the phase-matching condition, $\delta(x)$ is the Dirac delta function. As we can see from \eqref{qudit:freq}, the frequencies oscillate with peaks at $\omega_s -\omega_i=2\pi/\tau$, thus generating a frequency entangled qudit state.\\
\textit{Path} -- One of the first degrees of freedom to be exploited and manipulated for the generation of multilevel quantum systems has been the path. In Ref.~\cite{zukowski1997}, $\mathrm{\dot{Z}}$ukowski and co-authors showed how combinations of multiport beam-splitters (BS) can be suitably engineered to have non-classical correlations in higher dimension in path. Such a seminal study can be considered of great importance, especially for integrated photonic devices. Indeed, BSs implementation on a chip, either silicon or silica-based, is straightforward and this made it possible to develop integrated sources emitting entangled photon pairs up to dimension $d=15$, as we are going to see in the next subsection~\cite{Wang2018}. An innovative technique to generate entanglement in higher dimensions by using path has been presented in Ref.~\cite{Krennpath2017}. The authors show how two-photon arbitrary high-dimensional entanglement can be generated by path identity. In particular, starting with separable (non-entangled) photons, photon pairs are created in different crystals and their paths are overlapped, producing several types of entanglement in high-dimensions. The authors achieve proper control on \textit{which state} generate by using modes and phase shifters, showing the great flexibility of the method.\\
\textit{Degrees of freedom composition} -- Every degree of freedom analyzed offers practical advantages and constraints. Nonetheless, a combination of them sometimes can help to explore larger Hilbert spaces with fewer difficulties. For instance, although polarization bases live in a two-dimensional Hilbert space, they can be combined with almost all the other degrees of freedom. Indeed, noteworthy and fundamental results have been achieved by considering \textit{hybrid} high-dimensional states. 
\begin{figure*}[ht!]
\centering
\includegraphics[width=0.99\textwidth]{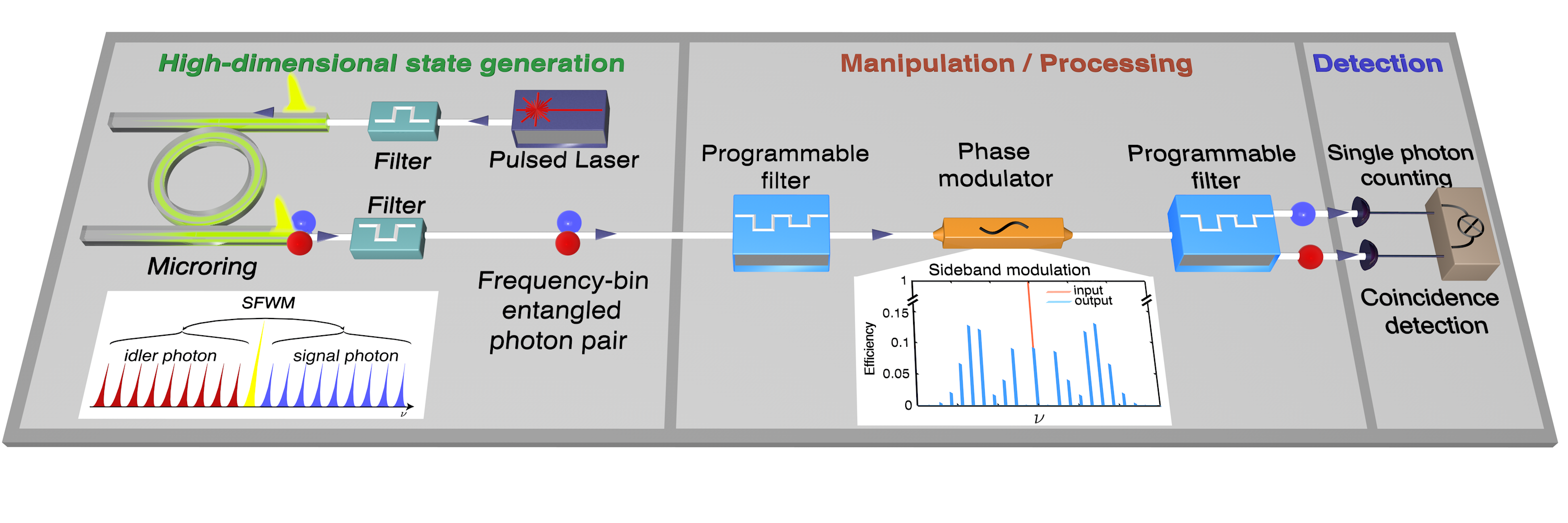}
\caption{{\bf Experimental setup of frequency high-dimensional chip.}
A mode-locked laser coupled to the integrated micro-ring excites precisely a single resonance. Spontaneous four-wave mixing (SFWM) (see left inset) process generates photon pairs (signal and idler) spectrally symmetric in a quantum superposition of the frequency modes. Programmable filters and a modulator were used for manipulating the quantum states, before the single-photon detector measurement. Reproduced with permission~\cite{Kues2017}. Copyright 2017, Nature.}
\label{fig:chip_kues}
\end{figure*}
Some examples of hybrid qudits are spin-orbit states (combination of polarization and OAM)~\cite{nagali2010experimental,vector2016,erhard2015,Cozzolino2019}, path-polarization states~\cite{Guo18,Ciampini19}, polarization-time states~\cite{Steinlechner2017} and frequency-path states~\cite{galmes2019}.

\subsection{Integrated platforms}
The route towards the full deployment of quantum technologies resides in the capacity of creating identical and replicable quantum devices.  
Integrated optical sources offer immense advantages due to their intrinsic scalability, high stability, and repeatable production process.
However, not all degrees of freedom can be efficiently manipulated on integrated circuits. For instance, the polarization of photons needs to be properly addressed in order for integrated chips to be able to carry two orthogonally polarized fields. Furthermore, the generation and propagation of OAM states through waveguides is very demanding as well, but small steps towards the reliable on-chip transmission and source integration of such states have been proved~\cite{chen2018mapping,chen2019vector}. As previously mentioned, frequency and path degrees of freedom are the two that can be more easily controlled and manipulated on integrated devices~\cite{maltese2019generation,matsuda2017generation,imany201850}. In the following, we are going to focus and discuss in detail two very important and impressive experiments, which use path and frequency encoded high-dimensional states respectively. The results obtained are fundamental to speed up the development of quantum technologies and to bridge the gap between the classical ones.\\
The first experiment involves path encoded qudits in silicon integrated platform~\cite{Wang2018}. 
Silicon quantum photonics is a promising candidate to further develop integrated quantum devices, as it offers intrinsic stability, high precision and integration with other classical devices.
The device allows the generation of high-dimensional entangled states with a controllable degree of entanglement. Figure~\ref{fig:chip_bristol} shows a schematic of the chip design. Photons entangled over $d$ spatial modes are generated by coherently pumping $d$ different single-photon sources. In particular, 16 photon sources, emitting photon pairs by spontaneous four-wave mixing process (SFWM), are integrated~\cite{Santagati2018}. Thus, an high-dimensional entangled state is created:
\begin{equation}
\ket{\Psi}_d=\sum_{k=0}^{d-1}c_k\ket{k}_s\ket{k}_i
\end{equation}  
where the qudit state $\ket{k}$ is associated to a photon in the $k$-th optical mode, the subscripts $s$ and $i$ stands for signal and idler and the coefficients $c_k$ are complex probability amplitudes. Such coefficients can be chosen arbitrarily by changing the pump distribution of the sources and the relative phase of the mode. This precise control is achieved with cascaded Mach-Zehnder interferometers (MZIs) at the input and phase shifters on each optical mode. By uniformly pumping the sources, maximally entangled states can be obtained. On the same device, linear optical circuits allow for the implementation of any local unitary transformation in $d$ dimensions. The authors estimate the indistinguishability of the 16 sources by performing a reverse Hong-Ou-Mandel interference and by calculating the visibility of the fringes. 
All the visibilities obtained are higher than 0.90, being higher than 0.98 in more than 80\% of the cases. 
Quantum state tomographies and certification on the system dimensionality as well as the violation of generalized Bell's inequalities (CGLMP) are performed. In addition, the authors studied unexplored quantum applications, which are quantum randomness expansion and self-testing on multidimensional states, thus showing exhaustively the potentials of such an integrated device.\\
The second experiment by M. Kues and colleagues~\cite{Kues2017} demonstrates the generation of high-dimensional frequency entangled states up to dimension $d=$10. 
The states are obtained by pumping a micro-ring resonator to provoke the SFWM process and hence generate pairs of photons in a superposition of multiple frequency modes. In particular, a spectrally filtered mode-locked laser excites a single resonance of the micro-ring, producing pairs of correlated signal and idler photons spectrally symmetric to the excitation field, as reported in Figure~\ref{fig:chip_kues}.
Thus, the quantum states are selected and manipulated using commercially available telecommunication programmable filters. The joint spectral intensity, describing the two-photon state’s frequency distribution, can therefore be measured, and Bell-test measurements and quantum state tomography can be carried out~\cite{Kues2017}. Also, the authors have sent a two-dimensional frequency-entangled state through a 24.2 km long fiber, and they prove the correct propagation of such states through Bell's inequality test.\\ 
In this section, we have reviewed all the possible platforms and schemes capable to generate high-dimensional states to our knowledge. We have divided them into two different classes: bulk and integrated platforms. The former constitutes the backbone of optics experiments and it is a very good approach for proof-of-principle experiments on quantum information and fundamental physics. However, due to its lack of scalability, it is not a good platform for advances in quantum technology, whereas the latter is more appropriate. Integrated optics limitations are mostly related to the degrees of freedom that can be exploited and properly controlled. Indeed, frequency and path are very well suited for integrated platforms, but devices able to manipulate and control with the same precision other degrees of freedom, \textit{e.g.} polarization and spatial modes, are still lacking. In terms of degrees of freedom, using the time to generate qudit states is a clever approach and it is also suitable for integrated devices. However, increasing dimensions by using time lowers the repetition rate of the generated states and this could be a non-trivial issue for technological applications. Finally, although the OAM of light constitutes a natural basis for high-dimensional states, it is very challenging to manipulate on integrated devices, thus it is mainly exploited in bulk optics experiments. Nonetheless, noteworthy results that might open the doors to integrated devices exploiting OAM have been achieved~\cite{cai2012integrated,chen2018mapping,chen2019vector,liu2018direct}.

\begin{figure*}[h!]
\centering
\includegraphics[width=0.99\textwidth]{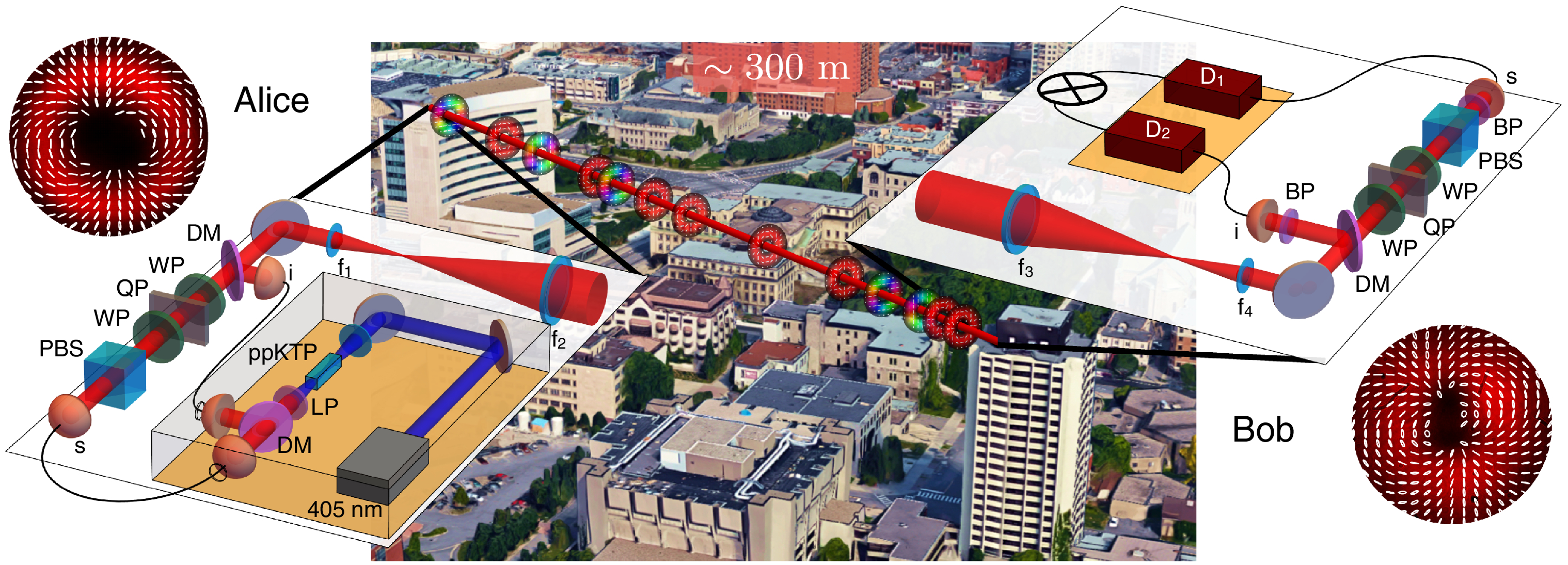}
\caption{{\bf Setup of the intracity OAM distribution.} Schematic of the transmitter (left) with a heralded single-photon source and state preparation. Alice prepares the quantum states using a polarizing beam splitter (PBS), wave plates (WP), and a q-plate (QP). The signal and idler photons are recombined on a dichroic mirror (DM) before the propagation on the free-space channel. Two telescopes comprised of lenses with focal lengths of $f_1=75$ mm, $f_2$ and $f_3$=400 mm (diameter of $75$ mm), and $f_4=50$ mm are used to enlarge and collect the beam, minimizing its divergence through the 0.3 km link. 
Bob, receiver (right), performs projective measurements on the quantum states and records the coincidences between signal and idler photons with detectors $D_1$ and $D_2$ and time-tagger unit. Examples of experimentally reconstructed polarization distributions of a structured mode using a continuous wave laser prepared by Alice (top left) and measured by Bob (bottom right) are shown in the insets. Legend: ppKTP, periodically poled KTP crystal; LP: long-pass filter; BP: band-pass filter. Map data: Google Maps, 2016. Reprinted figure with permission from Ref.~[140], The Optical Society.}
\label{fig:Ottawa}
\end{figure*}

\section{Quantum communications}
Subsequently the generation of high-dimensional quantum states, this section regards the propagation of such states through a communication channel, \textit{e.g.}, optical fiber, free-space or underwater links. Even though profound improvements have been made to generate and manipulate high dimensional quantum states, their reliable transmission, the cornerstone for future quantum networks, remains an open challenge.

\subsection{Free-space links}
The distribution of quantum states between distant
parties, connected by a free-space link, is one of the main technological challenges towards a global-scale quantum Internet.
Several proof-of-concept studies have already demonstrated the high-fidelity transmission of entangled photons up to 143 km for a ground link~\cite{Ma2012}, 1200 km with a satellite link for quantum communication on a global scale~\cite{Yin1140}, and transmission of attenuate laser up to Global Navigation Satellite Orbit~\cite{Dequal2016, Calderaro2018}.
However, until very recently, most of the demonstrations used a bipartite binary photonic system while only a few took advantage of qudit encoding. 
As described in the previous sections, the gain offered by high-dimensional systems can be applied to multiple areas.
In particular, for communication purposes, the ability to encode more information in a single-photon is a peculiar characteristic for pushing the entire field.\\
As introduced in the generation section, a straightforward way to generate high-dimensional quantum states is to use space encoding.
Spatial modes, \textit{e.g.} LG, can be adopted to implement high-dimensional quantum states without any constraint on the Hilbert space size. A first example is represented by the correct generation and detection of maximally entangled qutrits for quantum key distribution accomplished in an optical table~\cite{Groblacher2006}. S.~Gr\"{o}blacher \textit{et al.} used a parametric down-conversion scheme to generate pairs of photons entangled in their orbital angular momentum. The quantum key is then encoded in different LG modes created by tunable phase holograms.  The detection is accomplished with multiple beam splitters and holograms which allow projecting the quantum states in different bases.\\
Furthermore, thanks to their small divergence angle and intrinsic rotational symmetry, LG modes are suitable for long-distance free-space optical communication. However, there are practical limitations on the finite size of apertures in a realistic system, which limits the dimensions of the Hilbert space that can be used for communication; indeed, in free-space links, beam divergence must be taken into account~\cite{Nape2018}.\\ 
The main example of spatial encoded qudits over a free-space channel is represented by the work of A.~Sit and colleagues~\cite{Sit2017}, where they proved the correct transmission of LG modes (of dimension four) in a 300 meters intracity air link in Ottawa, as reported in Figure~\ref{fig:Ottawa}. 
\begin{figure*}[t]
\centering
\includegraphics[width=0.99\textwidth]{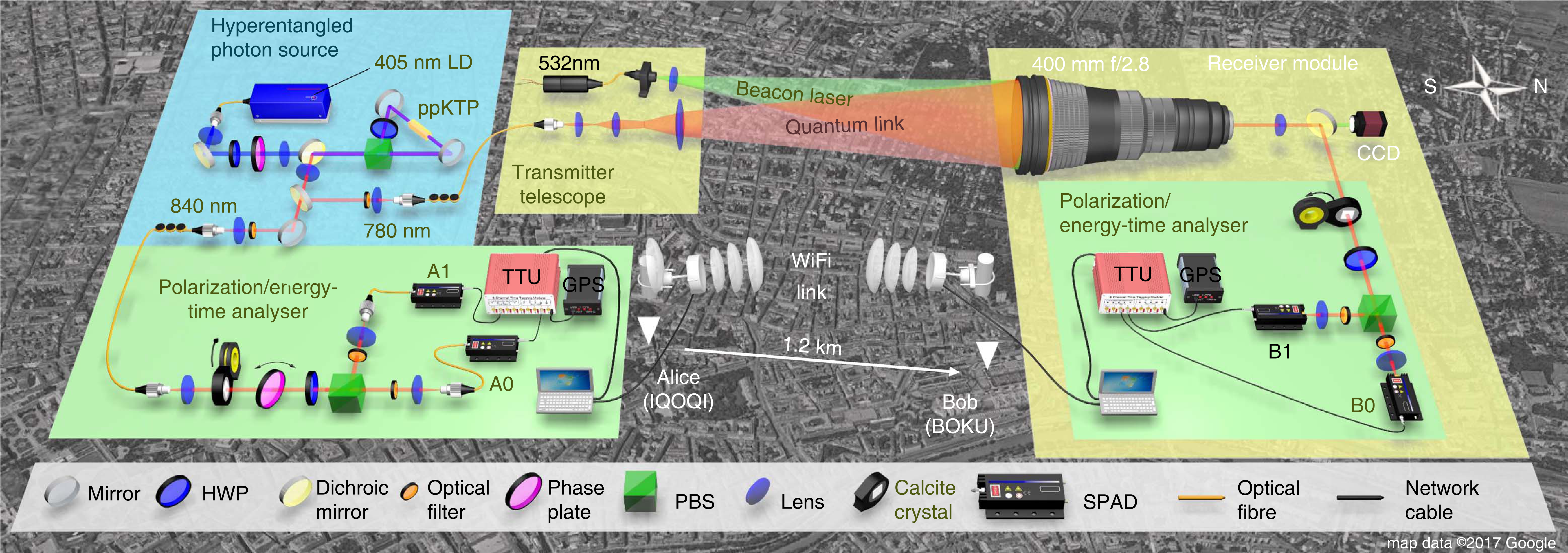}
\caption{{\bf Setup of the hyper-entangled distribution in Vienna.}
A hyper-entangled photon source was located in a laboratory at the IQOQI Vienna. The source utilized SPDC crystal, which was placed at the center of a Sagnac interferometer and pumped with a continuous-wave laser diode (LD) to obtain polarization/energy–time hyper-entangled photon pairs. Photon A was sent to Alice at IQOQI using a short fiber link, while photon B was guided to a transmitter telescope on the roof of the institute and sent to Bob at the BOKU via a 1.2-km-long free-space link. At Bob, the photons were collected using a large-aperture telephoto objective with a focal length of 400 mm. The beacon laser was separated from the hyper-entangled photons using a dichroic mirror and focused onto a CCD image sensor to maintain link alignment and to monitor atmospheric turbulence. Alice’s and Bob’s analyzer modules allowed for measurements in the polarization or energy–time basis.
Single-photon detection events were recorded with a GPS-disciplined time tagging unit (TTU) and stored on local hard drives for post-processing. Bob’s measurement data were streamed to Alice via a classical WiFi link to identify photon pairs in real time. Map data: Google Maps, 2017.
Reprinted with permission from Ref.~\cite{Steinlechner2017}, Nature Publishing group.}
\label{fig:vienna}
\end{figure*}
Their transmitter unit is composed of a parametric down-conversion single-photon source, where non-degenerate wavelengths are selected for the signal and idler photons.
The signal photon is subsequently used for key encoding by employing a q-plate combined with wave plates to prepare the ququarts in the mutually unbiased bases.
At the receiver side, the signal photon is projected in one of the states, while the idler photon is measured by a single-photon detector. Coincidence counts are registered and a key is extracted between Alice and Bob. 
To be noted that the receiver allows measuring only one of the quantum states at the time, limiting both the receiver efficiency and future applications. In other words, since Alice can prepare one of the four quantum states in each basis and Bob does not implement an optimal quantum receiver, the number of actual sifted bits is decreased by the probability of choosing the same symbol both for Alice and for Bob. This setup configuration ends up limiting the range of applications: some quantum protocols, such as complete device-independent demonstrations and loophole-free measurements for non-locality tests, require to measure all the possible outcomes at the same time. Indeed, for a $D$ dimensional Hilbert space detection loophole-free test, $D+1$ outcomes are required to strictly violate Bell's inequalities. However, projecting on $N<D$ outcomes, only a subset of all emitted pairs are measured, introducing possible classical correlation~\cite{Wang2018}.\\
Other examples, from the same authors, of high-dimensional protocols based on OAM modes are reported in~\cite{Bouchard2018experimental}, where many quantum protocols (BB84, Chau15, Singapore)\footnote{Chau15\cite{chau2015} is a new proposal for qubit-like qudit of protocols. In particular, it requires fewer resources, in terms of state preparation, compared to a full high-dimensional protocol; Singapore protocol instead, implements a specific POVM operator allowing a full tomography of the quantum states\cite{singapore2004}. 
} are studied in different dimensions, from 2 to 8, demonstrating an ideal range of application depending on the noise and on the system environment.\\
Besides LG modes, other spatial modes can be used for qudit encoding. S.~Etcheverry \textit{et al.}~\cite{Etcheverry2013} uses linear transverse momentum of weak coherent pulses as the degree of freedom for encoding a 16-dimensional qudit state. At Alice and Bob’s sites, the quantum states spanning the mutually unbiased bases are randomly produced with the help of a spatial light modulator, dynamically introducing relative phase shifts between the paths. The stability of the system, over a few meter link, is measured for several hours~\cite{Etcheverry2013}. Again, the receiver implemented in this experiment allows projection one quantum state at the time, penalizing the overall efficiency as discussed above.\\
\begin{figure*}[ht]
\centering
\includegraphics[width=0.99\textwidth]{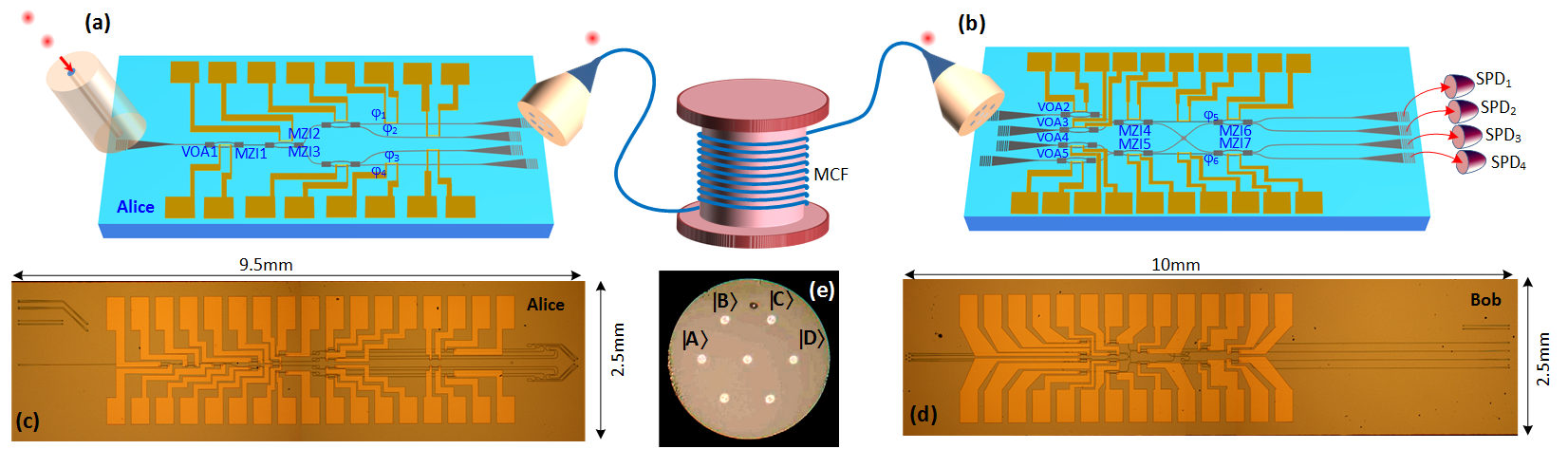}
\caption{{\bf Setup of the high-dimensional chip to chip experiment.} \textbf{(a)}, \textbf{(b)} Schematic of the integrated components for the high-dimensional QKD protocol
based on the multicore fiber (MCF). Labels are: variable optical attenuator (VOA); Mach-Zehnder interferometer (MZI), phases ($\varphi$); single photon detectors (SPD). \textbf{(c)},\textbf{ (d)} pictures of the integrated photonic chips used in the experiment. \textbf{(e)} Cross section of the MCF showing the four cores used.}
\label{fig:chip2chip}
\end{figure*}
Other degrees of freedom commonly adopted in free-space links are time-energy and polarization encoding. Both time-energy and polarization are not much affected by the effects of free-space propagation, \textit{i.e.} beam wandering and scintillation. In this direction F.~Steinlechner \textit{et al.} have demonstrated the correct propagation of high-dimensional entangled photons, exploiting hyper-entanglement between polarization and time-energy, over a 1.2 km free-space link in Vienna~\cite{Steinlechner2017}.
The source of hyper-entangled photons was based on the SPDC process using standard optical components and a long coherence time of the optical pump.
Both transmitter and receiver are equipped with a polarization analyzer module and an unbalanced polarization interferometer (based on calcite crystal) to convert the energy-time degree of freedom to polarization. Further details on the setup are reported in Figure~\ref{fig:vienna}.
The experimental data showed lower bound visibilities of 98\% and 91\% for polarization and energy-time respectively, corresponding to a minimum value of 0.94 and 0.77 \textit{ebits} (entangled bits) of entanglement of formation~\cite{Steinlechner2017}. Furthermore, by considering the combined hyper-entangled state, they obtained 1.46 of ebits of entanglement of formation~\cite{Steinlechner2017} and a Bell-state fidelity of 0.94, certifying the dimensionality of the system to $d=4$.\\
These experiments certify the capability of employing high-dimensional quantum states, encoded in multiple degrees of freedom, for free-space links. Thus, qudits could be useful for future quantum communication links, such as satellites to Earth connections and satellite to satellite communication.
\subsection{Fiber-based links}
Besides the transmission of high-dimensional quantum states in a free-space link, fiber links are the most attractive channel since the infrastructure is already in place and furthermore optical fiber communication is commonly used in our lives, \textit{e.g.} for the Internet backbone. Different fibers can be exploited for the propagation of high-dimensional quantum states: single-mode fibers (SMFs- most used and deployed), multimode fibers (MMFs- including few-mode fibers, and higher-order modes fibers) and multicore fibers (MCFs- special fibers with more than one core within the same cladding). Depending on the application, an optimal solution can be found in different fiber types. As an example, in data centers, where space is limited and a remarkably high number of connections are required, the footprint of the fibers is very important and hence the use of MCFs is a very attractive solution.\\
\textit{Single-mode fibers} -- Regarding the distribution of high-dimensional quantum states through SMFs, a first example is represented by high-dimensional time-bin encoding for high rate quantum key distribution protocols. N.~T. Islam~\textit{et al.} in Ref.~\cite{Islam2017}, demonstrated how high-dimensional quantum states can be used to generate a very high secret key rate (26 Mbit/s), in a one-way protocol with 4 dB channel loss (emulated with a variable attenuator) under the general coherent attacks scenario.\\
However, one-way protocols and technological imperfections of QKD devices open the possibility of successful eavesdropping methods, like side-channel attacks~\cite{Gerhardt2011}. Device-independent (DI) protocols or measurement-device-independent protocols (MDI) can overcome these limitations~\cite{Lo2011}. In this direction, two proof-of-concept of high-dimensional QKD in an MDI scheme has been proposed~\cite{Dellantonio2018} and proved~\cite{Islam2019}.\\
Another appropriate approach to distributing qudits through fibers is represented by a time-energy encoding. T. Zhong \textit{et al.} demonstrated in Ref.~\cite{Zhong2015} the correct propagation over a 20 km link of time-energy qudits up to dimension $d=$1024, proving 2.7 Mbit/s of secret key rate generation under the condition of collective Gaussian attack. Furthermore, T. Ikuta and H. Takesue reported in Ref.~\cite{Ikuta2018} the distribution of four-dimensional time-bin entangled quantum states between separated users located at 100 km distance.\\
As reported by these experiments, time-bin and time-energy encoding are very convenient ways to generate and propagate high-dimensional quantum states in SMFs- these degrees of freedom are stable throughout optical fiber transmission and require fairly simple setups- but present some limitations. High-dimensional time encoding has the main drawback of lowering the effective information rate, as given a fixed repetition rate at the transmitter the actual symbol rate decreases by increasing the dimension of the system.
Conversely, high-dimensional protocols based on different degrees of freedom do not exhibit this behavior, since a qudit time-duration results in being the same as that taken by a qubit. An example is given by space encoding exploiting multicore or multimode fibers, which will be addressed in the next subsections.\\
\textit{Multicore fibers} -- Multicore fibers present well-performant characteristics: they offer low losses (comparable with the standard single-mode fibers) and low cross-talk between cores (fundamental for reliable transmission of the qudits)~\cite{Saitoh2016}.
Previous experiments already demonstrated the capability of transferring spatial modes of light with high-fidelity up to a dimension equal to four~\cite{Ding2017,Lee2017,Canas2017}.\\
In particular, Y. Ding and coauthors used two silicon photonic platforms, connected by a 3 m MCF, for preparing and measuring the quantum states~\cite{Ding2017}, as reported in Figure~\ref{fig:chip2chip}. The qudits are path-encoded in the cores of the MCF and prepared by using integrated Mach-Zehnder interferometers and phase shifters, which allow creating at least two mutually unbiased bases necessary for a QKD protocol. Weak coherent pulses are injected into Alice's chip, and decoy-state technique is applied to avoid the photon number splitting attack. The QBER is measured to be below the coherent attacks limit for several minutes, proving the correct propagation of a four-dimensional quantum state over the MCF.
This work plays an important role in future quantum networks since it combines two very attractive solutions for the generation and the transmission of high-dimensional quantum states: silicon photonics and multicore fibers.
However, the main challenge in these fibers is to maintain the phase stability between different cores, required to preserve the coherence of the superposition states-- when the information is encoded in the relative phase between the cores the result is a long fiber interferometer. A possible solution is the use of phase-locked loop systems able to compensate for phase drifts in real-time, as proved in~\cite{Canas2017}, or the use of reference-frame independent protocols~\cite{Laing2010}.\\
\textit{Multimode fibers} -- A final approach is represented by the use of multimode fibers. Traditionally, fibers have been engineered to be single-mode, since this is generally advantageous in optical communication. Despite this, other interesting applications are enabled by multimode fibers~\cite{Ramachandran2008}.\\
For instance, recently D. Cozzolino and colleagues reported the first demonstration of a high-dimensional quantum state, encoded in a superposition of orbital angular momentum modes ($l=\pm$ 6, and $l=\pm 7$), transmitted over a 1.2 km air-core fiber~\cite{Cozzolino2018}. Weak coherent pulses are prepared in a four-dimensional Hilbert space by utilizing bulk and fiber optics. Qudits are then propagated through the OAM-carrying fiber and measured in two mutually unbiased bases. The measurement of the OAM states is realized by implementing a free-space OAM sorter followed by projective measurements, allowing simultaneous measurements of all the states within the same basis. Different QKD protocols are implemented to test the correct propagation of the quantum states.\\
Furthermore, H. Cao \textit{et al.} recently showed the correct propagation and detection of three-dimensional entangled states~\cite{Cao2018} encoded in the OAM degree of freedom. They used lower-order OAM modes ($l=0, \pm 1$) prepared with an SPDC and free-space optics to realize a three-dimensional entangled source with 88\% of fidelity. After the transmission over 1 km of fiber, the qudits are measured using an SLM and single-photon detectors. A fidelity measurement of 71\% and high-dimensional Bell inequalities violation proved the correct propagation of the qudit entangle states.\\
Other works have investigated these special fibers proving the transmission of hybrid vector vortex-polarization entanglement over an air-core fiber~\cite{Cozzolino2019}, the distribution of bidimensional structured photons in a vortex fiber~\cite{Sit2018} and the transmission of spatially encoded qudits over few meters of multimode fiber~\cite{Amitonova2018}.\\
Despite these proof-of-concept experiments, the propagation of qudits encoded in OAM through special fibers is still challenging. The main limitations are the phase instability between the modes and intermodal dispersion. Theoretical work, with a more appropriate design and simulation, is necessary to engineer new fibers with less intermodal dispersion and crosstalk, with values suitable for future quantum communications.\\
Another interesting application worth to be considered is the use of multimode fiber for programming linear quantum networks. Linear optical networks are good candidates for future realization of quantum computing. However, limitations in terms of scalability and performance arise from current implementations. S. Leedumrongwatthanakun and colleagues report in Ref.~\cite{Leedumrongwatthanakun2019} the implementation of a fully programmable high-dimensional linear optical network by using spatial and polarization mixing processes in a multimode fiber.

\subsection{Underwater links}
The experimental implementation of practical quantum communication systems has so far been mainly limited to fibers and free-space links. 
However, during the last few years, the community has started to investigate quantum communication in an underwater environment.\\
In 2012, M.~Lanzagorta proposed the idea of bringing the technology of free-space quantum communication in the water~\cite{lanzagorta2012} by performing a feasibility analysis on the BB84 protocol in point-to-point communication. During the last years, the paper was followed by other theoretical investigation~\cite{Xu2018,Tarantino2019}, in which not only point-to-point links were considered, but also non-line of sight underwater communication was studied. Besides, few experimental demonstrations investigated the propagation of polarization-based quantum states~\cite{Ji2017,Hu2018_under}.\\
However, optical communication in an aquatic environment is subjected to multiple degradation factors: high losses, strong turbulence effects and external noise (sun or moon radiation).
These factors can be directly translated to higher noise in the communication system, which will influence the final performances of the communication, limiting the total distance and the key rate. However, since qudits are intrinsically more robust to the noise they can be used for underwater channels.\\
Similarly to free-space links, qudits encoded in spatial modes are suitable for the generation and the transmission of large dimensional quantum states.
In this direction, a recent study from F.~Bouchard and colleagues~\cite{Bouchard2018} has investigated the effects of turbulence on an underwater quantum channel with high-dimensional quantum states encoded in spatial modes. Photon pairs are generated via SPDC and, by using a spatial light modulator, Alice prepared the quantum states to propagate over a 3 m link. Bob projected the quantum states into the different bases using an SLM, while single-photon detectors were used to measure coincidences.
To prove the correct propagation of the quantum states, three proof-of-concept QKD protocols have been demonstrated. A two-dimensional BB84 protocol, exploiting OAM modes, proved the correctness of the quantum states with a QBER around 6.57\%. A six-state protocol, instead, allowed a lower QBER of 6.35\% generating 0.395 bit per sifted photons. Qudits were also investigated (qutrits and ququarts) employing different OAM modes $l=0, \pm 1$, $\pm 2$. The results report a QBER of 11.73\% (below the threshold of 15.95\%) for the qutrit system and 29.77\% for the ququart case, which is above the 18.93\% threshold of the collective attacks. The errors are attributed to aberrations induced by the underwater turbulence introducing crosstalk between OAM modes. 
Since the oscillations introduced by the water turbulence are of the order of tens of Hertz, an adaptive optics system might be used to correct the wavefront~\cite{sorelli2019} and to decrease the inter-modal crosstalk.\\   
In these sections,  experimental demonstrations of high-dimensional quantum state transmission through different types of communication channels have been reported. Many of the experiments are proof of concept realizations and, compared to ordinary qubit implementations, they required a more advanced setup both for the transmitter and for the receiver, not always practical with the current technology. Also, the maximum dimensionality explored and successfully transported over a communication channel is four (ququart).
However, such limitations can be overcome by the use of integrated photonics, which allows exceptional control on the generation and manipulation of the quantum states, and by further improvement in the realization of multicore/multimode optical fibers. 
Summarizing, it is currently difficult to foresee an imminent use of high-dimensional quantum schemes in quantum networks, but depending on the channel characteristics (noise, distance, hardware availability) qudits can play a very important role in the future quantum systems.

\section{What's next: future perspectives and open questions}
The quantum internet represents the final goal of quantum communication. It can disclose a whole universe of new applications that can enlighten new fundamental physics questions or boost secure communication and remote quantum computing~\cite{Trabesinger2012,Svore2016}. The key ingredient of the quantum internet is the capability to distribute and store entanglement between separated users. Despite the big efforts over the last decade, long-distance transmission and long-time storage of entangled states remain open challenges. In this direction, high-dimensional quantum states can play a prominent role, due to their enhanced robustness to noise and higher information capacity. Nonetheless, their exploitation is not straightforward, mainly due to experimental limitations and theoretical problems still open.
Indeed, generalizing and experimentally proving protocols like entanglement swapping or quantum teleportation (primary tasks in quantum communication) by using qudits is not trivial~\cite{Calsamiglia2002}. Exploiting non-linear optics devices can be an approach~\cite{Vitali2000}, but the highly probabilistic processes involved can limit the advantages of the higher dimensionality. Ancillary photons or hyper-entanglement between two or more degrees of freedom can represent a solution to implement such protocols, as some proof-of-concept experiments have proven~\cite{Hu2019,Hueaat2018,Zhang2017,Guo2019}.
From a general perspective, the realization of a quantum network cannot happen regardless of quantum memories. The quantum community has already proved the capacity of storing multidimensional states in quantum memories, but very little has been done to prove the compatibility between external qudit sources with quantum memories for high-dimensional states~\cite{Ding_m_2014}. Hence, we think that in the next years more research needs to be done towards the conjunction of these two branches.\\
In terms of quantum foundations, an open problem related to high-dimensional entangled states is the certification of the actual entangled dimensions of a quantum system. In fact, despite the theoretical debates we have not addressed in this review, the experimental certification of high-dimensional entanglement requires full state tomographies of bipartite $d$-dimensional systems. This implies that $(d+1)^2$  global product bases measurements are needed, which quickly becomes impractical for high dimensions. Recently, it has been proposed by J. Bavaresco \textit{et al.}~\cite{Bavaresco2018} a new way to certify the dimensions of entangled qudits with carefully constructed measurements in two bases. This result will be of great advantage and will boost the research on entangled qudits towards unexplored applications.\\ 
As final a remark, we would like to raise a question which hitherto has not a conclusive answer:
\textit{how much information can be encoded into a single qudit?}
While in general quantum theory does not
impose any limit on the mass or dimension of a quantum system, a quantum-to-classical transition can be expected. Many groups are working to squeeze the boundaries between the microscopic and macroscopic description of the world, and thus trying to understand the limits, if there are, between quantum and classical states. A way to address these questions is represented by quantum optomechanics, which, by coupling mechanical oscillators to optical fields, not only can study quantum information applications but also it has great potential to test quantum physics at the microscopic edge~\cite{Marinkovi2018}. In this sense, qudits can be key players by helping in the boundary definition process between classical and quantum world, and thus bearing to a deeper understanding of the physical world.

\section{Conclusion}
During the last few years, an increasing number of reports on theoretical and experimental advances in the generation, propagation, and measurement of high-dimensional quantum states have been published. These include the generation of high-dimensional quantum states, up to dimension 15, in a silicon photonic platform~\cite{Wang2018}, the transmission of high-dimensional quantum states through multicore~\cite{Ding2017} and multimode fibers~\cite{Cozzolino2018} and the distribution of OAM modes~\cite{Sit2017} and hyper-entangled photons~\cite{Steinlechner2017} via a free-space intracity channel. These achievements are possible owing to technological progress and a better understanding of the physical principles underlying larger Hilbert spaces. 
We believe that high-dimensional quantum states will play a fundamental role in the next quantum technological leap.

\section*{Biography}
\vspace{0.5cm}
\begin{figure}[h!]
\centering
\includegraphics[width=0.12\textwidth]{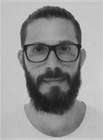}
\end{figure}

\textbf{Daniele Cozzolino} is a Ph.D. candidate of the SPOC center at the Department of Photonics Engineering at the Technical University of Denmark (DTU). He obtained his B.Sc and M.Sc in Physics at the University of Naples Federico II. 
His research interests are focused on quantum information and fundamental physics.

\begin{figure}[h!]
\centering
\includegraphics[width=0.12\textwidth]{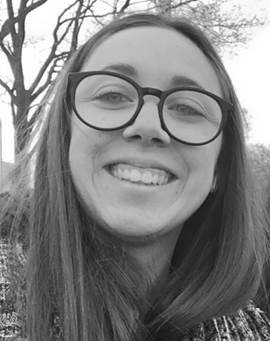}
\end{figure}
\textbf{Beatrice Da Lio} is a Ph.D. candidate of the SPOC center at the Department of Photonics Engineering at the Technical University of Denmark (DTU). She obtained her B.Sc at the University of Padova. She holds a double M.Sc degree in Engineering Telecommunication at the University of Padova and at the Technical University of Denmark. 
Her research interests are focused on quantum cryptography and quantum communication.

\begin{figure}[h!]
\centering
\includegraphics[width=0.12\textwidth]{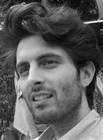}
\end{figure}
\textbf{Davide Bacco} is an Assistant Professor at the Department of Photonics Engineering at the Technical University of Denmark (DTU). He received his degree in Engineering Telecommunication in 2011 at the University of Padova, Italy. In 2015 he finished in the same University the Ph.D. degree on Science Technology and Spatial Measures (CISAS).
His research interests regard quantum communication and silicon photonics for quantum communications.

\begin{figure}[h!]
\centering
\includegraphics[width=0.12\textwidth]{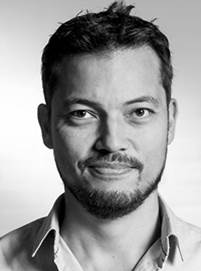}
\end{figure}
\textbf{Leif K. Oxenløwe} is the group leader of the High-Speed Optical Communications group at DTU Fotonik, at the Technical University of Denmark (DTU), and he is leader of the Centre of Excellence SPOC (Silicon Photonics for Optical Communications). He received his B.Sc. and M.Sc. degrees in physics and astronomy from the Niels Bohr Institute, University of Copenhagen, in 1996 and 1998, respectively. He received his Ph.D. degree in 2002 from DTU and since 2009 he is Professor of Photonic Communication Technologies. His research interests are focused on silicon photonics for optical processing and high-speed optical communication.

\section*{Funding Information}
This work is supported by the Centre of Excellence SPOC - Silicon Photonics for Optical Communications (ref DNRF123), by the People Programme (Marie Curie Actions) of the European Union's Seventh Framework Programme (FP7/2007-2013) under REA grant agreement n$^\circ$ $609405$ (COFUNDPostdocDTU).

\section*{Acknowledgments}
We thank Y. Ding, K. Rottwitt and M. Galili for their help and contribution to our works. We would like to show our gratitude to our collaborators from University of Bristol, Sapienza University of Rome, Boston University, University of Copenhagen and OFS Denmark.  

\section*{Conflict of interest}
The authors declare no conflict of interest.
\section*{Keywords}
high-dimensional, qudit, quantum communications


\end{document}